\documentclass[twocolumn,showpacs,showkeys,preprintnumbers,amsmath,amssymb,pre]{revtex4}

\usepackage{graphicx}
\usepackage{dcolumn}
\usepackage{bm}

\def\bfkone{\mbox{${\bm{k}_1}$}} 
\def\bfktwo{\mbox{${\bm{k}_2}$}} 
\def\bfkthr{\mbox{${\bm{k}_3}$}}

\def\bfsx{\mbox{$\scriptstyle{\bm{x}}$}} 
\def\bfskone{\mbox{$\scriptstyle{\bm{k}_1}$}} 
\def\bfsktwo{\mbox{$\scriptstyle{\bm{k}_2}$}} 
\def\bfskthr{\mbox{$\scriptstyle{\bm{k}_3}$}} 

\begin{document}

\preprint{APS/123-QED}

\title{Quasipatterns in parametrically forced systems}

\author{A. M. Rucklidge}%
 \affiliation{Department of Applied Mathematics,  
 University of Leeds, Leeds LS2 9JT, UK}%
\author{M. Silber}%
 \affiliation{Department of Engineering Sciences and Applied Mathematics,\\
              Northwestern University, Evanston, IL 60208, USA}%

\date{\today}

 \begin{abstract}
 We examine two mechanisms that have been put forward to explain the selection
of quasipatterns in single and multi-frequency forced Faraday wave experiments.
Both mechanisms can be used to generate stable quasipatterns in a
parametrically forced partial differential equation that shares some
characteristics of the Faraday wave experiment. One mechanism, which is robust
and works with single-frequency forcing, does not select a specific
quasipattern: we find, for two different forcing strengths, a 12-fold
quasipattern and the first known example of a spontaneously formed 14-fold
quasipattern. The second mechanism, which requires more delicate tuning, can be
used to select particular angles between wavevectors in the quasipattern.
 \end{abstract}


\maketitle

\section{Introduction}

The Faraday wave experiment consists of a horizontal layer of fluid
that spontaneously develops a pattern of standing waves on its surface as it is
driven by vertical oscillation with amplitude exceeding a critical value;
see \cite{Kudrolli1996a,Arbell2002} for surveys. Faraday wave experiments have
repeatedly produced new patterns of behaviour that required new ideas for their
explanation. An outstanding example of this was the discovery of {\it
quasipatterns} in experiments with one frequency~\cite{Christiansen1992} and
with two commensurate frequencies~\cite{Edwards1994}. Quasipatterns
do not have any translation symmetries, but their spatial Fourier transforms
have 8, 10 or 12-fold rotational order. There is as yet no satisfactory
mathematical treatment of quasipatterns~\cite{Rucklidge2003}.

Two mechanisms have been proposed for quasipattern formation, both building on
ideas of Newell and Pomeau~\cite{Newell1993}. One applies to single frequency
forcing~\cite{Zhang1996} and has been tested
experimentally~\cite{Westra2003}. Another was developed to
explain the origin of the two length scales in superlattice
patterns~\cite{Topaz2002,Porter2004} found in two-frequency
experiments~\cite{Kudrolli1998}. The ideas have not been tested quantitatively,
but have been used qualitatively to control
quasipattern~\cite{Arbell2002,Ding2006} and long-scale superlattice
pattern~\cite{Epstein2006} formation in two and three-frequency experiments.

With advances in computing power, we are able to go to larger domains, higher
resolutions and longer integration times to obtain very clean examples of
approximate quasipatterns, going further than previous numerical
studies~\cite{Zhang1998}. In addition, we report here the first example of a
spontaneously formed 14-fold quasipattern. One issue, which we do not address
here, is the distinction between a true quasipattern and an approximate
quasipattern, as found in numerical experiments with periodic boundary
conditions. We take the point of view that a system that produces an
approximate quasipattern is a good place to look for true quasipatterns.

One aim of this paper is to demonstrate that both proposed mechanisms for
quasipattern formation are viable. In order to claim convincingly that we
understand the pattern selection process, we have designed a partial
differential equation (PDE) and forcing functions that produce {\it a priori}
the required patterns. The PDE has multi-frequency forcing and shares many of
the characteristics of the real Faraday wave experiment, but has an easily
controllable dispersion relation and simple nonlinear terms:
 \begin{eqnarray} 
 \frac{\partial A}{\partial t} &=& (\mu+i\omega) A  
          + (\alpha+i\beta)\nabla^2 A  
          + (\gamma+i\delta)\nabla^4A  \nonumber\\
     & & {} + Q_1 A^2 + Q_2 |A|^2 + C |A|^2A 
          + i\hbox{Re}(A) f(t) 
 \label{eq:pde}
 \end{eqnarray} 
where $f(t)$ is a real-valued forcing function with period~$2\pi$, $A(x,y,t)$
is a complex-valued function, $\mu<0$, $\omega$, $\alpha$, $\beta$, $\gamma$
and $\delta$ are real parameters, and $Q_1$, $Q_2$, $C$ are complex parameters.
 
\section{Pattern selection} 
 
\begin{figure} 
\hbox to \hsize{\hfil 
 \hbox to 0.45\hsize{(a)\hfil}\hfil 
 \hbox to 0.45\hsize{(b)\hfil}\hfil}
\hbox to \hsize{\hfil 
  \mbox{\includegraphics[width=0.45\hsize]{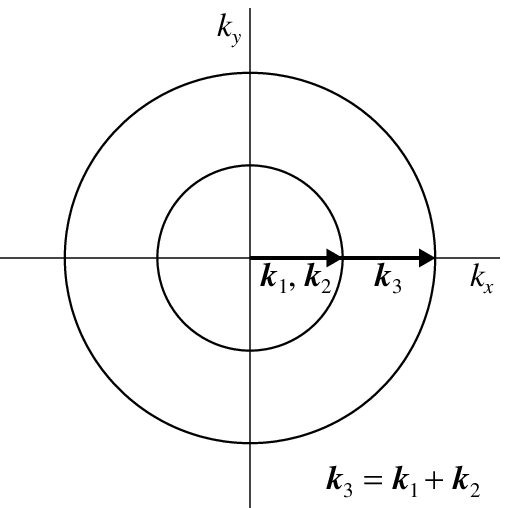}}\hfil 
  \mbox{\includegraphics[width=0.45\hsize]{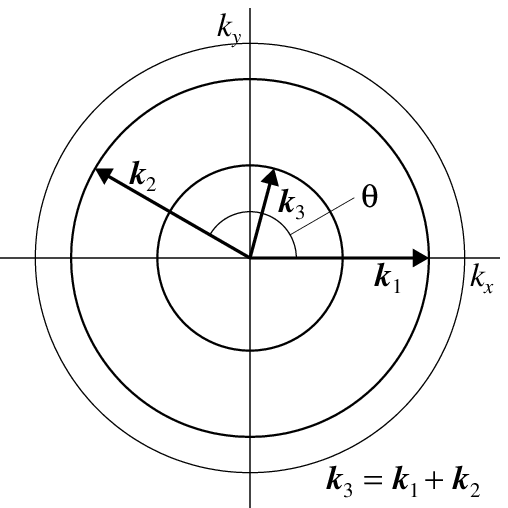}} 
\hfil} 
\caption{(a)~If the dispersion relation satisfies $\Omega(2k_m)=2\Omega(k_m)$,
then two modes with wavenumber~$k_m$ and aligned wavevectors $\bfkone=\bfktwo$
(inner circle) resonate in space and time with a mode with $\bfkthr=2\bfkone$
(outer circle). 
 (b)~With two-frequency forcing, consider two modes with wavevectors $\bfkone$
and $\bfktwo$, with the same wavenumber~$k_m$, and with $\Omega(k_m)=m/2$
(middle circle). The nonlinear combination of these two waves can, in the
presence of forcing at frequency~$n$, interact with a mode with
wavevector~$\bfkthr$ (inner circle), provided $\bfkthr=\bfkone+\bfktwo$ and
$\Omega(k_3)=|m-n|$.}
 \label{fig:resonances} 
 \end{figure}

Resonant triads play a key role in the understanding of pattern selection
mechanisms.  Consider a two (or more) frequency forcing function of the form
 \begin{equation}\label{eq:ft} 
 f(t)=f_m\cos(mt+\phi_m) + f_n\cos(nt+\phi_n) + ...,
 \end{equation} 
 where $m$ and $n$ are integers, $f_m$ and $f_n$ are amplitudes, and $\phi_m$
and $\phi_n$ are phases. We consider $m$ to be the dominant driving frequency,
and focus on a pair of waves, each with wavenumber $k_m$ satisfying the linear
dispersion relation $\Omega(k_m)=m/2$. These waves have the correct natural
frequency to be driven parametrically by the forcing~$f(t)$. We write the
critical modes in traveling wave form
 $z_1 e^{i\bfskone\cdot\bfsx+imt/2}$ and  
 $z_2 e^{i\bfsktwo\cdot\bfsx+imt/2}$.  
 These waves will interact nonlinearly with waves 
 $z_3 e^{i\bfskthr\cdot\bfsx+i\Omega(k_3) t}$, where  
 $\bfkthr=\bfkone+\bfktwo$ and $\Omega(k_3)$ is the frequency associated 
with $k_3$, provided that either (1)~the same resonance condition is met with 
the temporal frequencies, {\it i.e.}, 
 $\Omega(k_3) =\frac{m}{2}+\frac{m}{2}$, 
or (2)~any mismatch 
 $\Delta=|\Omega(k_3)-\frac{m}{2}-\frac{m}{2}|$ in this temporal resonance 
condition can be compensated by the forcing~$f(t)$. 
The first case corresponds to the $1:2$ 
resonance, which occurs even for single frequency forcing ($f_n=0$), and the 
second applies, {\it e.g.}, to two-frequency forcing  
with the third wave oscillating at 
the difference frequency: $\Omega(k_3)=|m-n|$ and $\Delta=n$. 
Note that in both 
cases, the temporal frequency $\Omega(k_3)$ determines the angle $\theta$ 
between the wave-vectors $\bfkone$ and $\bfktwo$ via the dispersion relation
(figure~\ref{fig:resonances}), 
and therefore provides a possible selection mechanism for certain preferred 
angles appearing in the power spectrum associated with the pattern. Selecting 
an angle of $0^\circ$ (figure~\ref{fig:resonances}a) is a special case. 

The nonlinear interactions of the modes can be understood by considering 
resonant triad equations describing standing wave patterns, which take the form 
 \begin{align}\label{eq:resonanttriad} 
 \dot{z}_1&= \lambda z_1+q_1 \bar{z_2}z_3+(a|z_1|^2+b|z_2|^2)z_1+\cdots\nonumber\\ 
 \dot{z}_2&= \lambda z_2+q_1 \bar{z_1}z_3+(a|z_2|^2+b|z_1|^2)z_2+\cdots\\ 
 \dot{z}_3&= \lambda_3 z_3+q_3z_1z_2+\cdots,\nonumber 
 \end{align} 
where all coefficients are real, and the dot refers to timescales long compared
to the forcing period. The quadratic coupling coefficients $q_j$ are {\rm O}(1)
in the forcing in the $1:2$ resonance case, and {\rm O}$(|f_n|)$ in the
difference frequency case. For other angles $\theta$ between the wavevectors
$\bfkone$ and $\bfktwo$ we expect $q_1\approx q_3\approx 0$ because the
temporal resonance condition for the triad of waves is not met. Here we are
assuming that the $z_3$-mode is damped when $\lambda$ goes through zero ({\it
i.e.}, $\lambda_3<0$ in~(\ref{eq:resonanttriad})), so $z_3$~can be eliminated
via center manifold reduction near the bifurcation point ($z_3\approx
\frac{q_3z_1z_2}{|\lambda_3|}$), resulting in the bifurcation problem
 \begin{align}\label{eq:ampsrhombs}
 \dot{z}_1&= \lambda z_1-(|z_1|^2+B_{\theta}|z_2|^2)z_1\nonumber\\ 
 \dot{z}_2&= \lambda z_2-(|z_2|^2+B_{\theta}|z_1|^2)z_2\ , 
 \end{align} 
where we have rescaled $z_1$ and $z_2$ by a factor of $1/\sqrt{|a|}$ and 
assumed that $a<0$. Here $B_{\theta}= b/a+\frac{q_1q_3}{a|\lambda_3|}$
includes the contribution from the slaved mode~$z_3$, and depends
on the angle~$\theta$ between the two wavevectors $\bfkone$ and~$\bfktwo$.

The function $B_{\theta}$ has important consequences for the stability of
regular patterns. Within the context of~(\ref{eq:ampsrhombs}), stripes are
stable if $B_{\theta}>1$, while rhombs associated with a given angle~$\theta$
are preferred if $|B_{\theta}|<1$. By judicious choice of forcing frequencies,
we have some ability to control the magnitude of $B_\theta$ over a range of
angles~$\theta$~\cite{Porter2004}, which allows the enhancement or suppression
of certain combinations of wavevectors in the resulting patterns.
Alternatively, if we choose forcing frequencies that select an angle of
$0^\circ$, then this can lead to a large resonant contribution: $a$ can become
large~\cite{Zhang1996}. This causes the rescaled
cross-coupling coefficient $B_{\theta}$ to be small over a broad range of
$\theta$ away from $\theta=0$.  (As $\theta\to 0$, it can be shown that
$B_\theta\to 2$.)

\begin{figure} 
\hbox to \hsize{(a)\hfil}
\hbox to \hsize{\hfil 
  \mbox{\includegraphics[width=0.99\hsize]{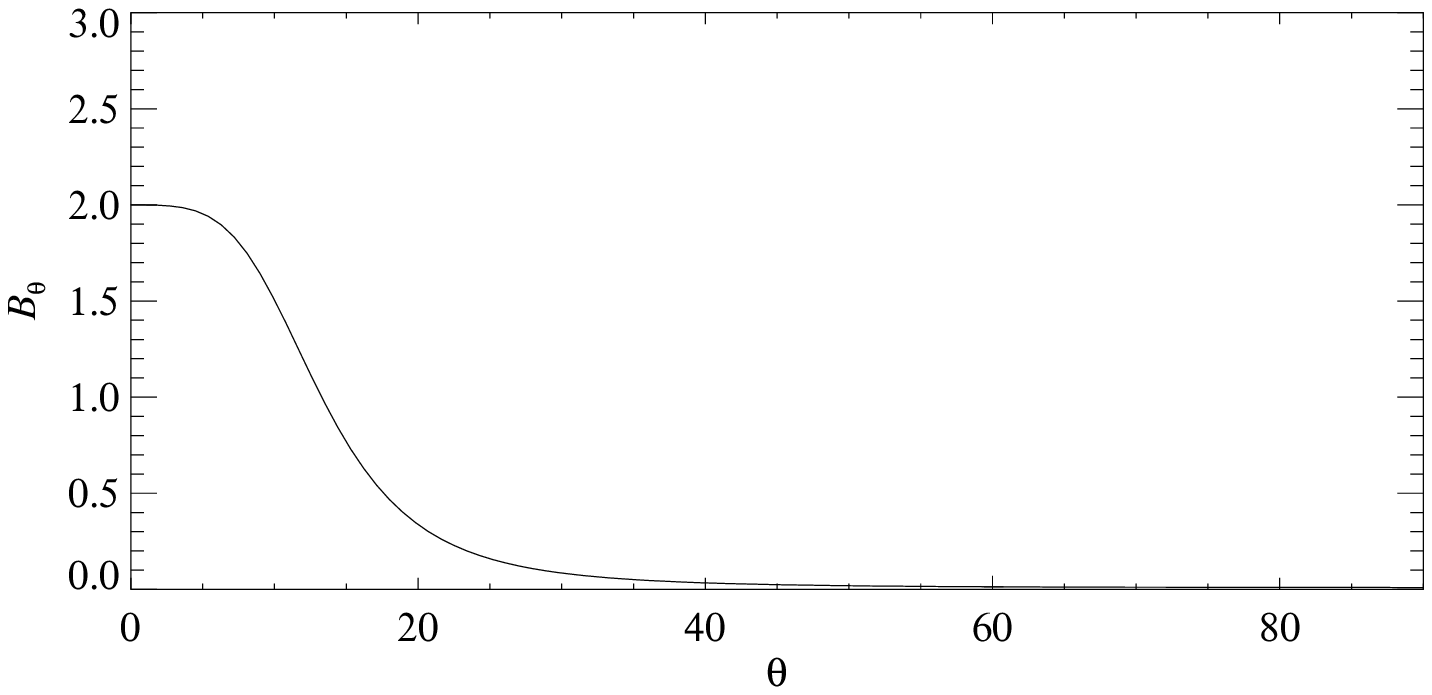}} 
\hfil} 
\hbox to \hsize{(b)\hfil}
\hbox to \hsize{\hfil 
  \mbox{\includegraphics[width=0.99\hsize]{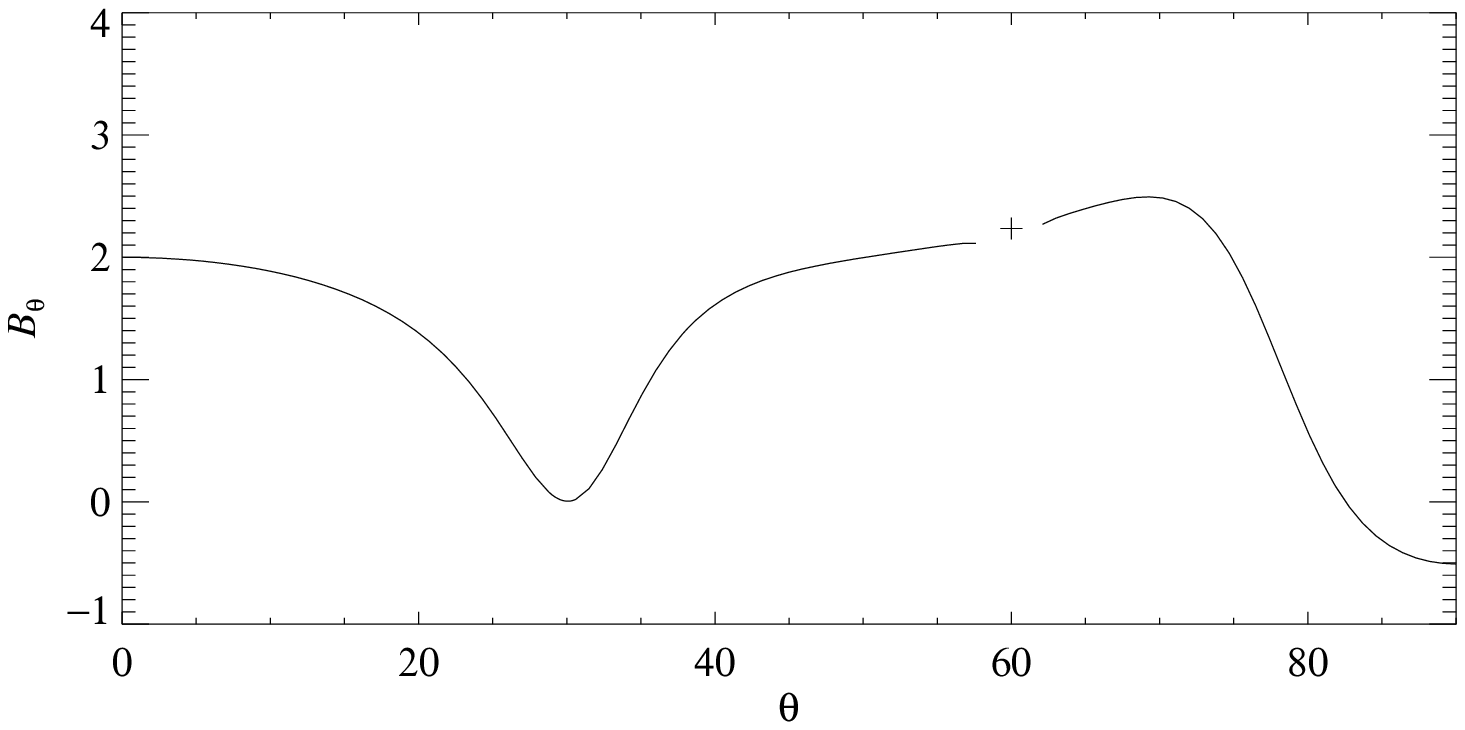}} 
\hfil} 
\caption{$B_{\theta}$ for the two cases. (a)~single frequency forcing with 
$1:2$ resonance. The parameter values are $\omega=\frac{1}{3}$, 
$\beta=-\frac{1}{6}$, $\delta=0$, $\mu=-0.005$, $\alpha=0.001$, $\gamma=0$,
$Q_1=3+4i$, $Q_2=-6+8i$, $C=-1+10i$, $m=1$, $\phi_1=0$ and $f_1=0.024002$.
(b)~multi-frequency $(4,5,8)$ forcing, with $\omega=0.633975$,
$\beta=-1.366025$, $\delta=0$, $\mu=-0.2$, $\alpha=-0.2$, $\gamma=-0.15$,
$Q_1=1+i$, $Q_2=-2+2i$, $C=-1+10i$, $f_4=0.53437$, $f_5=0.76316$,
$f_8=1.49063$, $\phi_4=0$, $\phi_5=0$ and $\phi_8=0$. The $+$ symbols are the 
result of a separate calculation.}
\label{fig:BthetaQP} 
\end{figure} 
 
\section{Results} 

We present parameter values that demonstrate that the two mechanisms are viable 
methods of predicting parameter values for stable approximate quasipatterns.

\begin{figure*} 
\hbox to \hsize{\hfil 
 \hbox to 0.38\hsize{\hfil (a)\hfil}\hfil 
 \hbox to 0.38\hsize{\hfil (b)\hfil}\hfil} 
\vspace{1ex} 
\hbox to \hsize{\hfil 
  \mbox{\includegraphics[width=0.38\hsize]{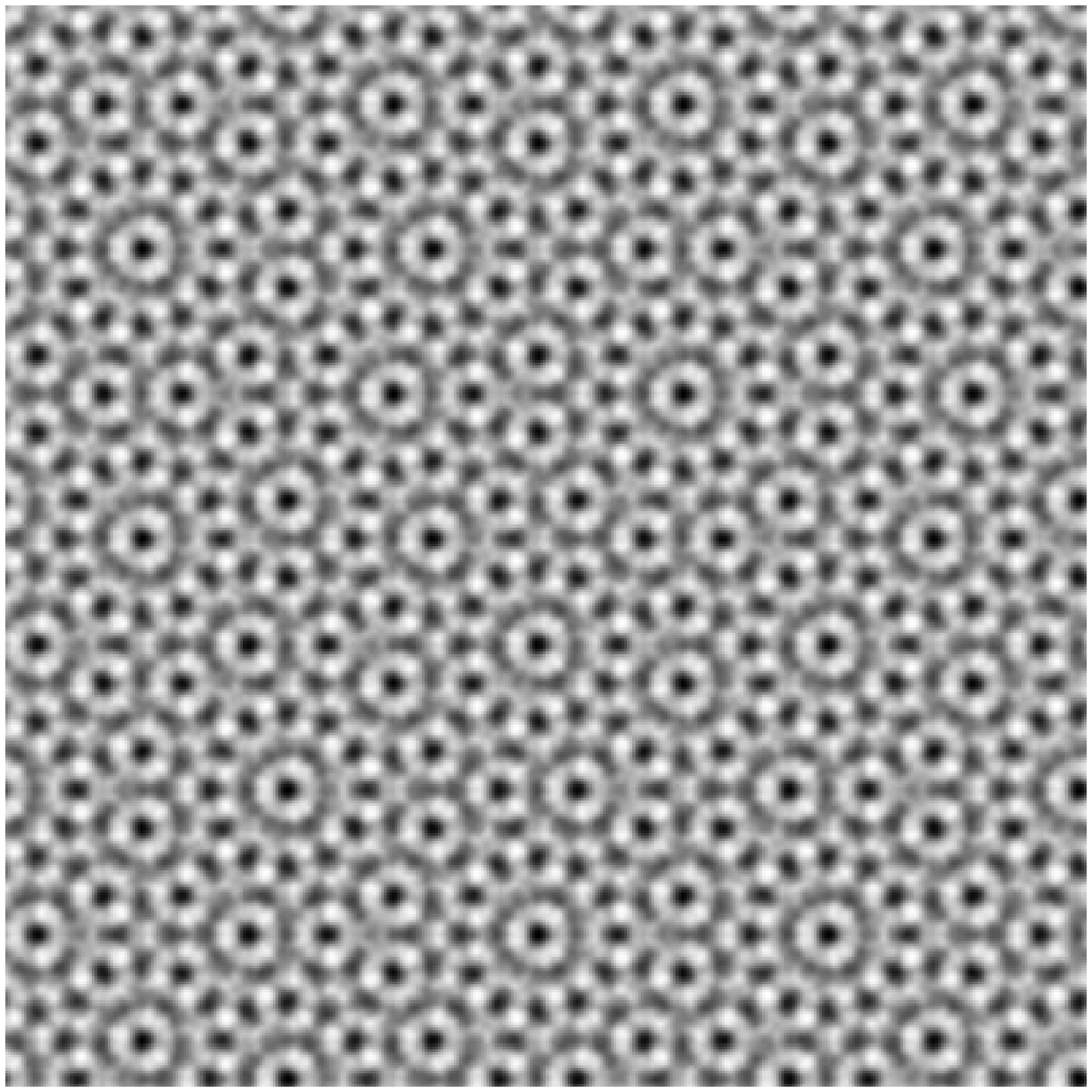}}\hfil 
  \mbox{\includegraphics[width=0.38\hsize]{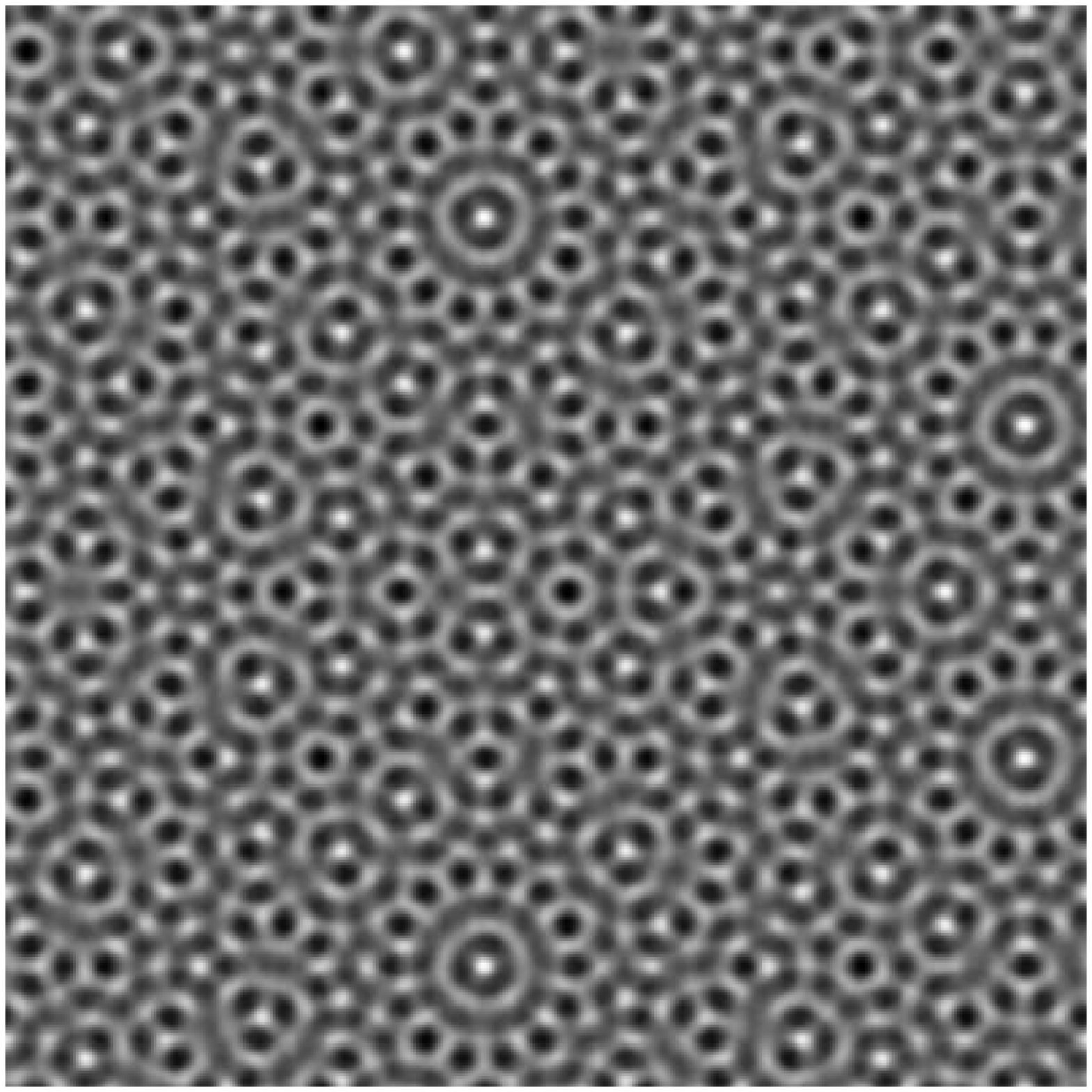}}\hfil} 
\vspace{1ex} 
\hbox to \hsize{\hfil 
 \hbox to 0.84\hsize{\hfil (c)\hfil}\hfil} 
\vspace{1ex} 
\hbox to \hsize{\hfil 
  \mbox{\includegraphics[width=0.84\hsize]{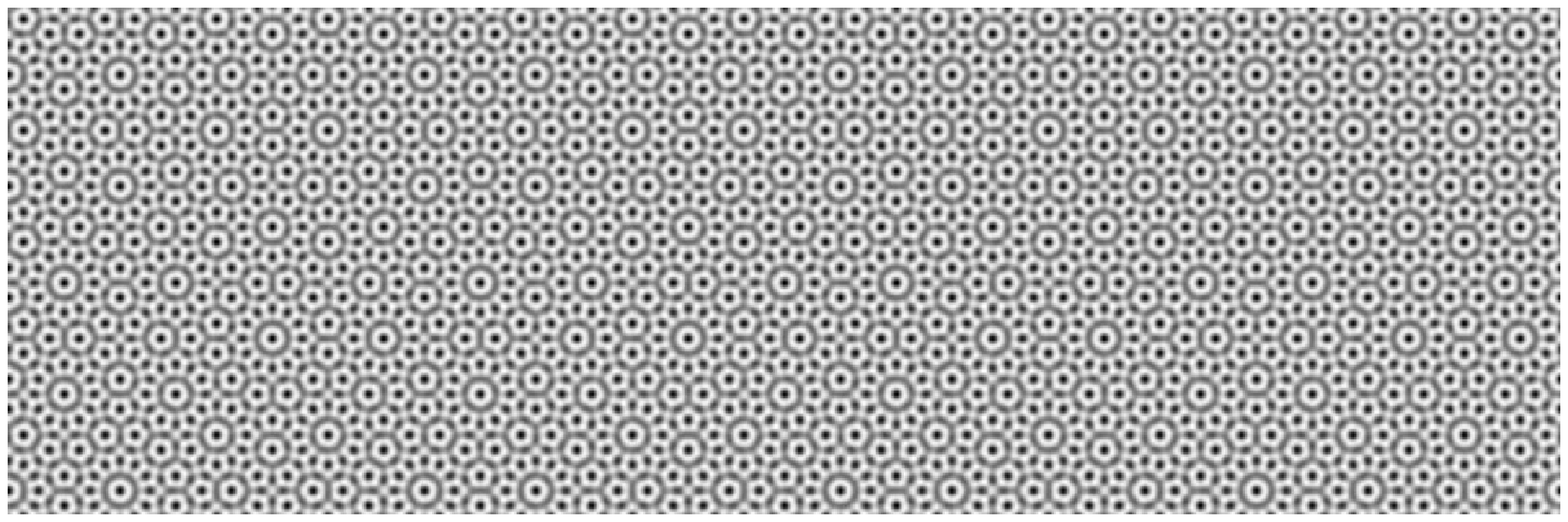}}\hfil}
\caption{(a)~With parameter values as in figure~\ref{fig:BthetaQP}a, in a 
domain $30\times30$ wavelengths, and forced at $1.1$ times the critical 
amplitude, we find subharmonic 12-fold quasipatterns. (b)~At $1.3$ times 
critical, the 12-fold quasipattern is unstable and is replaced by a 14-fold 
quasipattern.
c)~With parameter values as 
in figure~\ref{fig:BthetaQP}b and with $(f_4,f_5,f_8)$ set at $1.003$ times 
their critical values, we find harmonic 12-fold quasipatterns in a 
$112\times112$ domain (only a third is shown).}
\label{fig:QPexamples} 
\end{figure*} 


The dispersion relation of the PDE~(\ref{eq:pde}) is $\Omega(k)=\omega-\beta
k^2 + \delta k^4$. With single-frequency forcing, we choose $m=1$, 
and a spatial scale so that modes with $k=1$ are driven subharmonically:
$\Omega(1)=\frac{1}{2}$. To have $1:2$ resonance in space and time, we impose
$\Omega(2)=1$, which leads to $\omega=\frac{1}{3}+4\delta$ and
$\beta=-\frac{1}{6}+5\delta$. We choose $\delta=0$, small values for the
damping coefficients~$\mu$, $\alpha$ and~$\gamma$, and order~one values for the
nonlinear coefficients. We solve the linear stability problem to find the
critical value of the amplitude~$f_1$ in the forcing function, and use weakly
nonlinear theory to calculate $B_{\theta}$ (figure~\ref{fig:BthetaQP}a). This
curve has $B_0=2$, but $B_{\theta}$ drops away sharply, and is close to zero
for $\theta\geq30^\circ$, for the reasons explained above. We use $B_{\theta}$
at $30^\circ$, $60^\circ$ and $90^\circ$ and find that, within the restrictions
of a twelve-mode expansion, 12-fold quasipatterns are stable.

A numerical solution of the PDE~(\ref{eq:pde}) at $1.1$ times critical is shown
in figure~\ref{fig:QPexamples}(a), in a square domain with periodic boundary
conditions, of size $30\times30$ wavelengths, with $512^2$ Fourier modes
(dealiased). The timestepping method was the fourth-order
ETDRK4~\cite{Cox2002}, with 20 timesteps per period of the forcing. The
solution is an approximate quasipattern: the primary modes that make up the
pattern are $(30,0)$ and $(26,15)$ and their reflections, in units of basic
lattice vectors. These two wavevectors are $29.98^\circ$ apart, and differ in
length by $0.05\%$. The amplitudes of the modes differ by $0.5\%$. The initial
condition was not in any invariant subspace, and the PDE was integrated for
$160\,000$ periods of the forcing. However, when we increase the forcing to
$1.3$~times critical, we find that the 12-fold quasipattern is unstable and
is replaced (after a transient of $50\,000$ periods) by a 14-fold quasipattern
(figure~\ref{fig:QPexamples}b). In this case, the modes are $(30,0)$,
$(27,13)$, $(19,23)$ and $(7,29)$, differing in length by $0.5\%$ and having
angles within $1.5^\circ$ of $360^\circ/14$. The amplitudes differ by about
$10\%$.


The second method of producing quasipatterns involves the weakly damped
difference frequency mode, and is more selective, but also requires some
fine-tuning of the parameters. In order to use triad interactions to encourage
modes at $30^\circ$, we choose $m=4$, $n=5$ forcing, setting $\Omega(1)=2$, and
requiring that a wavenumber involved in $30^\circ$ mode interactions
($k^2=2-\sqrt{3}$) correspond to the difference frequency: $\Omega(k)=1$. One
solution is $\omega=0.633975$, $\beta=-1.366025$ and $\delta=0$. Twelve-fold
quasipatterns also require modes at $90^\circ$ to be favoured, and for these
choices of parameters, $\Omega(\sqrt{2})$ is~$3.37$. Although this is not
particularly close to~$4$, we can use $1:2$ resonance (driving at frequency~8)
to control the $90^\circ$ interaction. The resulting $B_{\theta}$ curve 
(figure~\ref{fig:BthetaQP}b) shows pronounced dips at $30^\circ$ and
$90^\circ$ as required. Again, $B_{30}$, $B_{60}$ and $B_{90}$ are used to show
that, within a 12-amplitude cubic truncation, 12-fold quasipatterns are stable, 
this time between $0.9995$ and $1.0095$ times critical. Squares are also stable
above $1.0015$ times critical.

A numerical solution of the PDE~(\ref{eq:pde}) at $1.003$ times critical is
shown in figure~\ref{fig:QPexamples}(c), in a periodic domain $112\times112$
wavelengths (integrated using $1536^2$ Fourier modes). This solution was
followed for over $10\,000$ forcing periods. The important wavevectors are
$(112,0)$ and $(97,56)$, which are $29.9987^\circ$ apart and differ in length
by $0.004\%$. The amplitudes of these modes differ by~$1\%$.

\section{Discussion}

We have investigated two mechanisms for quasipattern formation for
Faraday waves within a single PDE model of pattern formation via parametric
forcing, and have demonstrated that both mechanisms are viable. One uses $1:2$
resonance in space and time to enhance the self-interaction coefficient~$a$ and
so suppress the cross-coupling coefficient $B_{\theta}$ for angles greater than
about~$30^\circ$, and leads to ``turbulent crystals''~\cite{Newell1993}. Within
this framework, there is little distinction between 8, 10, 12 or 14-fold
quasipatterns, or indeed any other combination of modes, and there is no way of
knowing in advance which pattern will be formed. The mechanism is robust (the
patterns are found well above onset), and requires only a single frequency in
the forcing. A dispersion relation that supports $1:2$ resonance in space and
time is needed.

The existence of 14-fold (and higher) quasipatterns has been suggested
before~\cite{Zhang1996,Rucklidge2003,Topaz2004}, but we have presented here the
first example of a spontaneously formed 14-fold quasipattern that is a stable
solution of a parametrically forced PDE. Examples where 14-fold symmetry is
imposed externally have been reported in optical
experiments~\cite{Pampaloni1995}. The Fourier spectra of 12-fold and 14-fold
quasipatterns are very different~\cite{Rucklidge2003}, which may have
consequences for their mathematical treatment.

The second mechanism uses three-wave interactions involving a damped mode
associated with the difference of the two frequencies in the forcing to select
a particular angle ($30^\circ$ in the example presented here). Using different
primary frequencies, or altering the dispersion relation, allows other angles,
or combinations of angles, to be selected. The advantage is that a forcing
function can be designed to produce a particular pattern. On the other hand,
the strongest control of~$B_{\theta}$ occurs for parameters close to the
bicritical point, which limits the range of validity of the weakly nonlinear
theory used to compute stability. This issue will be pursued elsewhere.

It should be noted that the stability calculations done here are in the
framework of a twelve-mode amplitude expansion truncated a cubic order. The 
fact that stable 12-fold quasipatterns are found where they are expected
demonstrates that this approach provides useful information about the
stability of the approximate quasipatterns, in spite of the concerns about
small divisors raised by~\cite{Rucklidge2003}.


 We are grateful for support from the EPSRC (GR/S45928/01) and to the NSF
(DMS-0309667). We are also grateful to many people who have helped shape these
ideas: Jay Fineberg, Ian Melbourne, Michael Proctor, Jeff Porter, Anne Skeldon
and Jorge Vi\~nals. Finally, we acknowledge the hospitality of the Isaac Newton
Institute for Mathematical Sciences.


\bibliography{rs_pre}

\end{document}